\input harvmac
\def\ZZ{\hbox{Z\kern-.4emZ}}
\def\RR{\hbox{R\kern-.6emR}}

\lref\BredbergPV{
  I.~Bredberg, T.~Hartman, W.~Song and A.~Strominger,
  ``Black Hole Superradiance From Kerr/CFT,''
  arXiv:0907.3477 [hep-th].
}

\lref\MisnerKX{
  C.~W.~Misner,
  ``Interpretation of gravitational-wave observations,''
  Phys.\ Rev.\ Lett.\  {\bf 28}, 994 (1972).
}
\lref\TeukolskyHA{
  S.~A.~Teukolsky,
  ``Perturbations of a rotating black hole. 1. Fundamental equations for
  gravitational electromagnetic and neutrino field perturbations,''
  Astrophys.\ J.\  {\bf 185}, 635 (1973).
}

\lref\PressZZ{
  W.~H.~Press and S.~A.~Teukolsky,
  ``Perturbations of a Rotating Black Hole. II. Dynamical Stability of the Kerr
  Metric,''
  Astrophys.\ J.\  {\bf 185}, 649 (1973).
}

\lref\FrolovJH{
  V.~P.~Frolov and K.~S.~Thorne,
  Phys.\ Rev.\  D {\bf 39}, 2125 (1989).
}
\lref\CveticUW{
  M.~Cveti\v c and F.~Larsen,
  ``General rotating black holes in string theory: Greybody factors and  event
  horizons,''
  Phys.\ Rev.\  D {\bf 56}, 4994 (1997)
  [arXiv:hep-th/9705192].
}

\lref\CveticXV{
  M.~Cveti\v c and F.~Larsen,
  ``Greybody factors for rotating black holes in four dimensions,''
  Nucl.\ Phys.\  B {\bf 506}, 107 (1997)
  [arXiv:hep-th/9706071].
}

\lref\CveticVP{
  M.~Cveti\v c and F.~Larsen,
  ``Black hole horizons and the thermodynamics of strings,''
  Nucl.\ Phys.\ Proc.\ Suppl.\  {\bf 62}, 443 (1998)
  [Nucl.\ Phys.\ Proc.\ Suppl.\  {\bf 68}, 55 (1998)]
  [arXiv:hep-th/9708090].
}
\lref\CveticAP{
  M.~Cveti\v c and F.~Larsen,
  ``Greybody factors for black holes in four dimensions: Particles with
  spin,''
  Phys.\ Rev.\  D {\bf 57}, 6297 (1998)
  [arXiv:hep-th/9712118].
}

\lref\LarsenGE{
  F.~Larsen,
  ``A string model of black hole microstates,''
  Phys.\ Rev.\  D {\bf 56}, 1005 (1997)
  [arXiv:hep-th/9702153].
}

\lref\DiasNJ{
  O.~J.~C.~Dias, R.~Emparan and A.~Maccarrone,
  ``Microscopic Theory of Black Hole Superradiance,''
  Phys.\ Rev.\  D {\bf 77}, 064018 (2008)
  [arXiv:0712.0791 [hep-th]].
}

\lref\BredbergPV{
  I.~Bredberg, T.~Hartman, W.~Song and A.~Strominger,
  ``Black Hole Superradiance From Kerr/CFT,''
  arXiv:0907.3477 [hep-th].
}

\lref\StromingerSH{
  A.~Strominger and C.~Vafa,
  ``Microscopic Origin of the Bekenstein-Hawking Entropy,''
  Phys.\ Lett.\  B {\bf 379}, 99 (1996)
  [arXiv:hep-th/9601029].
}

\lref\CastroMS{
  A.~Castro, D.~Grumiller, F.~Larsen and R.~McNees,
  ``Holographic Description of AdS$_2$ Black Holes,''
  JHEP {\bf 0811}, 052 (2008)
  [arXiv:0809.4264 [hep-th]];
  A.~Castro and F.~Larsen,
  ``Near Extremal Kerr Entropy from AdS_2 Quantum Gravity,''
  arXiv:0908.1121 [hep-th].
}
\lref\HartmanDQ{
  T.~Hartman and A.~Strominger,
  ``Central Charge for AdS$_2$ Quantum Gravity,''
  JHEP {\bf 0904}, 026 (2009)
  [arXiv:0803.3621 [hep-th]].
}

\lref\GuicaMU{
  M.~Guica, T.~Hartman, W.~Song and A.~Strominger,
  ``The Kerr/CFT Correspondence,''
  arXiv:0809.4266 [hep-th].
}

\lref\SenQY{
  A.~Sen,
  ``Black Hole Entropy Function, Attractors and Precision Counting of
  Microstates,''
  Gen.\ Rel.\ Grav.\  {\bf 40}, 2249 (2008)
  [arXiv:0708.1270 [hep-th]].
}

\lref\BardeenPX{
  J.~M.~Bardeen and G.~T.~Horowitz,
  ``The extreme Kerr throat geometry: A vacuum analog of AdS(2) x S(2),''
  Phys.\ Rev.\  D {\bf 60}, 104030 (1999)
  [arXiv:hep-th/9905099].
}

\lref\BalasubramanianKQ{
  V.~Balasubramanian, A.~Naqvi and J.~Simon,
  ``A multi-boundary AdS orbifold and DLCQ holography: A universal  holographic
  description of extremal black hole horizons,''
  JHEP {\bf 0408}, 023 (2004)
  [arXiv:hep-th/0311237].
}
\lref\BalasubramanianBG{
  V.~Balasubramanian, J.~de Boer, M.~M.~Sheikh-Jabbari and J.~Simon,
  ``What is a chiral 2d CFT? And what does it have to do with extremal black
  holes?,''
  arXiv:0906.3272 [hep-th].
}

\lref\BarnichBF{
  G.~Barnich and G.~Compere,
  ``Surface charge algebra in gauge theories and thermodynamic integrability,''
  J.\ Math.\ Phys.\  {\bf 49}, 042901 (2008)
  [arXiv:0708.2378 [gr-qc]].
}
\lref\CompereAZ{
  G.~Compere,
  ``Symmetries and conservation laws in Lagrangian gauge theories with
  applications to the mechanics of black holes and to gravity in three
  dimensions,''
  arXiv:0708.3153 [hep-th].
}
\lref\BarnichKQ{
  G.~Barnich and G.~Compere,
  ``Conserved charges and thermodynamics of the spinning Goedel black hole,''
  Phys.\ Rev.\ Lett.\  {\bf 95}, 031302 (2005)
  [arXiv:hep-th/0501102].
}
\lref\BanadosDA{
  M.~Banados, G.~Barnich, G.~Compere and A.~Gomberoff,
  ``Three dimensional origin of Goedel spacetimes and black holes,''
  Phys.\ Rev.\  D {\bf 73}, 044006 (2006)
  [arXiv:hep-th/0512105].
}

\lref\EmparanEN{
  R.~Emparan and A.~Maccarrone,
  ``Statistical Description of Rotating Kaluza-Klein Black Holes,''
  Phys.\ Rev.\  D {\bf 75}, 084006 (2007)
  [arXiv:hep-th/0701150].
}

\lref\DiasNJ{
  O.~J.~C.~Dias, R.~Emparan and A.~Maccarrone,
  ``Microscopic Theory of Black Hole Superradiance,''
  Phys.\ Rev.\  D {\bf 77}, 064018 (2008)
  [arXiv:0712.0791 [hep-th]].
}

\lref\BardeenPX{
  J.~M.~Bardeen and G.~T.~Horowitz,
  ``The extreme Kerr throat geometry: A vacuum analog of AdS(2) x S(2),''
  Phys.\ Rev.\  D {\bf 60}, 104030 (1999)
  [arXiv:hep-th/9905099].
}

\lref\BalasubramanianRE{
  V.~Balasubramanian and P.~Kraus,
  ``A stress tensor for anti-de Sitter gravity,''
  Commun.\ Math.\ Phys.\  {\bf 208}, 413 (1999)
  [arXiv:hep-th/9902121].
}

\lref\SkenderisWP{
  K.~Skenderis,
  ``Lecture notes on holographic renormalization,''
  Class.\ Quant.\ Grav.\  {\bf 19}, 5849 (2002)
  [arXiv:hep-th/0209067].
}

\lref\AmselEV{
  A.~J.~Amsel, G.~T.~Horowitz, D.~Marolf and M.~M.~Roberts,
  ``No Dynamics in the Extremal Kerr Throat,''
  arXiv:0906.2376 [hep-th].
}
\lref\DiasEX{
  O.~J.~C.~Dias, H.~S.~Reall and J.~E.~Santos,
  ``Kerr-CFT and gravitational perturbations,''
  arXiv:0906.2380 [hep-th].
}

\lref\MaldacenaBW{
  J.~M.~Maldacena and A.~Strominger,
  ``AdS(3) black holes and a stringy exclusion principle,''
  JHEP {\bf 9812}, 005 (1998)
  [arXiv:hep-th/9804085].
}

\lref\DijkgraafFQ{
  R.~Dijkgraaf, J.~M.~Maldacena, G.~W.~Moore and E.~P.~Verlinde,
  ``A black hole farey tail,''
  arXiv:hep-th/0005003.
}

\lref\KrausVZ{
  P.~Kraus and F.~Larsen,
  ``Microscopic Black Hole Entropy in Theories with Higher Derivatives,''
  JHEP {\bf 0509}, 034 (2005)
  [arXiv:hep-th/0506176].
}

\lref\LarsenBU{
  F.~Larsen,
  ``Anti-de Sitter spaces and nonextreme black holes,''
  arXiv:hep-th/9806071.
}

\lref\CveticXV{
  M.~Cveti\v c and F.~Larsen,
  ``Greybody factors for rotating black holes in four dimensions,''
  Nucl.\ Phys.\  B {\bf 506}, 107 (1997)
  [arXiv:hep-th/9706071].
}

\lref\MaldacenaDS{
  J.~M.~Maldacena and L.~Susskind,
  ``D-branes and Fat Black Holes,''
  Nucl.\ Phys.\  B {\bf 475}, 679 (1996)
  [arXiv:hep-th/9604042].
}
\lref\KunduriVF{
  H.~K.~Kunduri, J.~Lucietti and H.~S.~Reall,
  ``Near-horizon symmetries of extremal black holes,''
  Class.\ Quant.\ Grav.\  {\bf 24}, 4169 (2007)
  [arXiv:0705.4214 [hep-th]].
}

\lref\StromingerYG{
  A.~Strominger,
  ``AdS(2) quantum gravity and string theory,''
  JHEP {\bf 9901}, 007 (1999)
  [arXiv:hep-th/9809027].
}
\lref\ChoFZ{
  J.~H.~Cho, T.~Lee and G.~W.~Semenoff,
  ``Two dimensional anti-de Sitter space and discrete light cone
  quantization,''
  Phys.\ Lett.\  B {\bf 468}, 52 (1999)
  [arXiv:hep-th/9906078].
}

\lref\GimonUR{
  E.~G.~Gimon and P.~Horava,
  ``Astrophysical Violations of the Kerr Bound as a Possible Signature of
  String Theory,''
  Phys.\ Lett.\  B {\bf 672}, 299 (2009)
  [arXiv:0706.2873 [hep-th]].
}
\lref\AzeyanagiBJ{
  T.~Azeyanagi, T.~Nishioka and T.~Takayanagi,
  ``Near Extremal Black Hole Entropy as Entanglement Entropy via AdS2/CFT1,''
  Phys.\ Rev.\  D {\bf 77}, 064005 (2008)
  [arXiv:0710.2956 [hep-th]].
}

\lref\GuptaKI{
  R.~K.~Gupta and A.~Sen,
  ``Ads(3)/CFT(2) to Ads(2)/CFT(1),''
  JHEP {\bf 0904}, 034 (2009)
  [arXiv:0806.0053 [hep-th]].
}

\lref\MaldacenaUZ{
  J.~M.~Maldacena, J.~Michelson and A.~Strominger,
  ``Anti-de Sitter fragmentation,''
  JHEP {\bf 9902}, 011 (1999)
  [arXiv:hep-th/9812073].
}

\lref\MaldacenaIH{
  J.~M.~Maldacena and A.~Strominger,
  ``Universal low-energy dynamics for rotating black holes,''
  Phys.\ Rev.\  D {\bf 56}, 4975 (1997)
  [arXiv:hep-th/9702015].
}
\lref\MathurET{
  S.~D.~Mathur,
  ``Absorption of angular momentum by black holes and D-branes,''
  Nucl.\ Phys.\  B {\bf 514}, 204 (1998)
  [arXiv:hep-th/9704156].
}
\lref\GubserQR{
  S.~S.~Gubser,
  ``Can the effective string see higher partial waves?,''
  Phys.\ Rev.\  D {\bf 56}, 4984 (1997)
  [arXiv:hep-th/9704195].
}

\lref\LarsenXM{
  F.~Larsen,
  ``The attractor mechanism in five dimensions,''
  Lect.\ Notes Phys.\  {\bf 755}, 249 (2008)
  [arXiv:hep-th/0608191].
}
\lref\CveticXZ{
  M.~Cveti\v c and D.~Youm,
  ``General Rotating Five Dimensional Black Holes of Toroidally Compactified
  Heterotic String,''
  Nucl.\ Phys.\  B {\bf 476}, 118 (1996)
  [arXiv:hep-th/9603100].
}

\lref\CveticKV{
  M.~Cveti\v c and D.~Youm,
  ``Entropy of Non-Extreme Charged Rotating Black Holes in String Theory,''
  Phys.\ Rev.\  D {\bf 54}, 2612 (1996)
  [arXiv:hep-th/9603147].
}

\lref\HartmanPB{
  T.~Hartman, K.~Murata, T.~Nishioka and A.~Strominger,
  ``CFT Duals for Extreme Black Holes,''
  JHEP {\bf 0904}, 019 (2009)
  [arXiv:0811.4393 [hep-th]].
}

\lref\MaldacenaIX{
  J.~M.~Maldacena and A.~Strominger,
  ``Black hole greybody factors and D-brane spectroscopy,''
  Phys.\ Rev.\  D {\bf 55}, 861 (1997)
  [arXiv:hep-th/9609026].
}

\lref\MaldacenaIH{
  J.~M.~Maldacena and A.~Strominger,
  ``Universal low-energy dynamics for rotating black holes,''
  Phys.\ Rev.\  D {\bf 56}, 4975 (1997)
  [arXiv:hep-th/9702015].
}
\lref\CallanTV{
  C.~G.~.~Callan, S.~S.~Gubser, I.~R.~Klebanov and A.~A.~Tseytlin,
  ``Absorption of fixed scalars and the D-brane approach to black holes,''
  Nucl.\ Phys.\  B {\bf 489}, 65 (1997)
  [arXiv:hep-th/9610172].
}
\lref\GubserZP{
  S.~S.~Gubser and I.~R.~Klebanov,
  ``Four-dimensional greybody factors and the effective string,''
  Phys.\ Rev.\ Lett.\  {\bf 77}, 4491 (1996)
  [arXiv:hep-th/9609076].
}
\lref\ChowDP{
  D.~D.~K.~Chow, M.~Cveti\v c, H.~Lu and C.~N.~Pope,
  ``Extremal Black Hole/CFT Correspondence in (Gauged) Supergravities,''
  arXiv:0812.2918 [hep-th].
}

\lref\LopesCardosoKY{
  G.~Lopes Cardoso, A.~Ceresole, G.~Dall'Agata, J.~M.~Oberreuter and J.~Perz,
 ``First-order flow equations for extremal black holes in very special
 geometry,''
  JHEP {\bf 0710}, 063 (2007)
  [arXiv:0706.3373 [hep-th]].
}

\lref\HottaWZ{
  K.~Hotta and T.~Kubota,
  ``Exact Solutions and the Attractor Mechanism in Non-BPS Black Holes,''
  Prog.\ Theor.\ Phys.\  {\bf 118}, 969 (2007)
  [arXiv:0707.4554 [hep-th]].
}
\lref\GimonGK{
  E.~G.~Gimon, F.~Larsen and J.~Simon,
  ``Black Holes in Supergravity: the non-BPS Branch,''
  JHEP {\bf 0801}, 040 (2008)
  [arXiv:0710.4967 [hep-th]].
  ``Constituent Model of Extremal non-BPS Black Holes,''
  JHEP {\bf 0907}, 052 (2009)
  [arXiv:0903.0719 [hep-th]].
  }
\lref\BenaEV{
  I.~Bena, G.~Dall'Agata, S.~Giusto, C.~Ruef and N.~P.~Warner,
  ``Non-BPS Black Rings and Black Holes in Taub-NUT,''
  JHEP {\bf 0906}, 015 (2009)
  [arXiv:0902.4526 [hep-th]].
}
\lref\deBoerIP{
  J.~de Boer,
  ``Six-dimensional supergravity on S**3 x AdS(3) and 2d conformal field
  theory,''
  Nucl.\ Phys.\  B {\bf 548}, 139 (1999)
  [arXiv:hep-th/9806104].
}

\lref\KastorGT{
  D.~Kastor and J.~H.~Traschen,
  ``A very effective string model?,''
  Phys.\ Rev.\  D {\bf 57}, 4862 (1998)
  [arXiv:hep-th/9707157].
}

\lref\DasWN{
  S.~R.~Das and S.~D.~Mathur,
  ``Comparing decay rates for black holes and D-branes,''
  Nucl.\ Phys.\  B {\bf 478}, 561 (1996)
  [arXiv:hep-th/9606185].
}

\lref\PeetES{
  A.~W.~Peet,
  ``The Bekenstein formula and string theory (N-brane theory),''
  Class.\ Quant.\ Grav.\  {\bf 15}, 3291 (1998)
  [arXiv:hep-th/9712253].
}
\lref\DavidWN{
  J.~R.~David, G.~Mandal and S.~R.~Wadia,
  ``Microscopic formulation of black holes in string theory,''
  Phys.\ Rept.\  {\bf 369}, 549 (2002)
  [arXiv:hep-th/0203048].
}
\lref\SenQY{
  A.~Sen,
  ``Black Hole Entropy Function, Attractors and Precision Counting of
  Microstates,''
  Gen.\ Rel.\ Grav.\  {\bf 40}, 2249 (2008)
  [arXiv:0708.1270 [hep-th]].
}

\lref\PiolinePF{
  B.~Pioline and J.~Troost,
  ``Schwinger pair production in AdS(2),''
  JHEP {\bf 0503}, 043 (2005)
  [arXiv:hep-th/0501169].
}
\lref\KimXV{
  S.~P.~Kim and D.~N.~Page,
  ``Schwinger Pair Production in $dS_2$ and $AdS_2$,''
  Phys.\ Rev.\  D {\bf 78}, 103517 (2008)
  [arXiv:0803.2555 [hep-th]].
}
\lref\AzeyanagiDK{
  T.~Azeyanagi, N.~Ogawa and S.~Terashima,
  ``The Kerr/CFT Correspondence and String Theory,''
  Phys.\ Rev.\  D {\bf 79}, 106009 (2009)
  [arXiv:0812.4883 [hep-th]].
}

\lref\BarnichJY{
  G.~Barnich and F.~Brandt,
  ``Covariant theory of asymptotic symmetries, conservation laws and  central
  charges,''
  Nucl.\ Phys.\  B {\bf 633}, 3 (2002)
  [arXiv:hep-th/0111246].
}
\lref\CompereIN{
  G.~Compere and S.~Detournay,
  ``Centrally extended symmetry algebra of asymptotically Goedel spacetimes,''
  JHEP {\bf 0703}, 098 (2007)
  [arXiv:hep-th/0701039].
}
\lref\CveticJA{
  M.~Cveti\v c and F.~Larsen,
  ``Statistical entropy of four-dimensional rotating black holes from
  near-horizon geometry,''
  Phys.\ Rev.\ Lett.\  {\bf 82}, 484 (1999)
  [arXiv:hep-th/9805146].
}
\lref\CveticXH{
  M.~Cveti\v c and F.~Larsen,
  ``Near horizon geometry of rotating black holes in five dimensions,''
  Nucl.\ Phys.\  B {\bf 531}, 239 (1998)
  [arXiv:hep-th/9805097].
}

\Title{\vbox{\baselineskip12pt 
\hbox{UPR-1210-T} 
\vskip-.5in}
}
{\vbox{\centerline{Greybody Factors and Charges in Kerr/CFT}
}}
\medskip
\centerline{\it
Mirjam Cveti\v c${}^{1}$ and Finn Larsen${}^{2}$
}
\bigskip
\centerline{${}^1$Department of Physics and Astronomy, University of Pennsylvania, Philadelphia, PA-19104, USA.}
\smallskip
\centerline{${}^2$Michigan Center for Theoretical Physics, 450 Church St., Ann Arbor,
MI-48109, USA.}
\smallskip

\vglue .3cm
\bigskip\bigskip\bigskip
\centerline{\bf Abstract}
\noindent
We compute greybody factors for near extreme Kerr black holes in $D=4$
and $D=5$. In $D=4$ we include four charges so that our solutions can be 
continuously deformed to the BPS limit. In $D=5$ we include two independent 
angular momenta so Left-Right symmetry is incorporated. We discuss the CFT 
interpretation of our emission amplitudes, including the overall frequency 
dependence and the dependence on all black hole parameters. We find that
all additional parameters can be incorporated Kerr/CFT, with central charge 
independent of $U(1)$ charges.

\Date{}

\newsec{Introduction}
The Hawking temperature of a black hole vanishes in the extreme limit $T_H\to 0$. 
It is therefore natural to interpret 
extreme black holes as ground states of the corresponding
quantum theory, and so presumably the simplest starting point for the analysis
of more general black holes with finite temperature. 
There are actually several inequivalent extreme limits. Writing the (inverse) 
Hawking temperature as 
\eqn\xa{
{1\over T_H} = {1\over 2} ({1\over T_R} + {1\over T_L})~,
}
we can take $T_H\to 0$ as:

\indent a) {\bf BPS}: $T_R\to 0$ {\it with supersymmetry preserved in the limit}. These are the
BPS black holes, the examples most analyzed in string theory (some reviews
are \refs{\PeetES,\DavidWN,\SenQY}).

\indent b) {\bf non-BPS}: $T_L\to 0$. This is the alternative extremal limit that breaks supersymmetry 
completely. It is the ``non-BPS branch'' that has been developed recently 
(including \refs{\LopesCardosoKY,\HottaWZ,\GimonGK,\BenaEV}). 

Recently there has been much progress on the description of extreme Kerr black holes
(including \refs{\GuicaMU,\HartmanPB,\BredbergPV}). 
One of the motivations for this particular extreme limit is that many astrophysical
black holes naturally spin up in the process of accretion, and so tend to approach the
extreme Kerr limit. The string theory description of extreme Kerr could therefore be
relevant for observations \refs{\GimonUR,\GuicaMU}. Importantly, the extreme Kerr limit 
defines a class distinct from those above:

\indent c) {\bf Extreme Kerr}: $T_R\to 0$ 
{\it with supersymmetry broken in the limit}, due to the
presence of angular momentum. 

The BPS black hole and the extreme Kerr black hole 
both correspond to a definite state in the $R$-sector, 
as far as classical considerations are concerned. The difference is that the BPS
black holes represent the true ground states, while the extreme Kerr black holes correspond
to states that have a condensate of angular momentum carriers (see eg.\DiasNJ). 
The condensate 
breaks supersymmetry and carries a macroscopic angular momentum; but it does 
not carry any macroscopic entropy and so $T_R\to 0$ just as in the true ground state
describing BPS black holes. 

An illuminating way to analyze the various limits is to compute the frequency dependent 
absorption cross-section of the black holes or, equivalently (due to detailed balance), 
the spectrum of Hawking 
emission \refs{\MaldacenaIX,\GubserZP,\CallanTV,\MaldacenaIH,\CveticUW,\CveticXV}

. This means solving the (massless) Klein-Gordon 
equation for a scalar field in the black hole background.
Despite the generality of our setting, the equation takes a strikingly simple
form:  it comprises some ``asymptotic'' terms and some ``near horizon'' terms. 
In {\it all} cases where the two groups of terms can be taken into account sequentially, 
the full solution takes the same form as the two-point correlator in a 2D CFT. 

One situation where the matching procedure is justified is for near extremal black holes 
where the two thermal scales $T_{L,R}$ establish a hierarchy. Significantly, 
the Left/Right structure described above shows that {\it the CFT underlying Kerr can be related 
by continuous deformation to the BPS black holes} that are well understood. All 
that is needed is that one must maintain the hierarchy $T_R\ll T_L$ as the angular 
momentum is turned off by tuning charges. 
This situation makes it interesting to keep all the 
charges as the Kerr limit is approached. This is one of the gaps in the literature that
we fill with this paper.  

The discussion so far was for 4D but there is a similar story for 5D black holes. In 5D
the R and L sectors are isomorphic, so there is no analogue of the non-BPS branch b). 
However, the relation between the BPS and the Kerr branches remains the same. Moreover,
since there are two angular momenta $J_{L,R}$, the near extreme limit defined by large 
$J_R$ generally leaves $J_L$ free. In this paper we take the dependence of this 
parameter into account.

The central charge of Kerr/CFT was determined in \GuicaMU\ using the method 
of \refs{\BarnichJY,\BarnichKQ,\CompereIN,\BarnichBF,\CompereAZ} to
determine the asymptotic symmetries. The result was later generalized to Kerr black holes
with one  and more charges in various dimensions \refs{\HartmanPB,\ChowDP,\AzeyanagiDK}. 
Here, we are interested in the striking CFT interpretation of the supergravity correlation functions. 
Two point functions do not immediately depend on the central charge, but they do 
depend on the complex structure of the space that the CFT is defined on. We discuss 
the relevant scales for the background with general charges.


The feature of the wave equation that leads to correlation functions reminiscent of a CFT 
is the hypergeometric nature of the near horizon regime. The hypergeometric function 
is a character of $SL(2,\RR)$, so one can try to interpret this structure as a remnant of a 
Virasoro algebra. On the other hand, the asymptotic terms corresponds to just the 
Coulomb-type gravitational scattering, which presumably does not probe the internal 
structure of the black holes. The hypergeometric nature of the near horizon equation 
remains for completely general black holes, with no extremality assumed. It is tempting 
to interpret this feature as a $SL(2,\RR)$ symmetry as well, albeit one that is broken 
by coupling to the asymptotic space. This could signal the presence of a Virasoro 
algebra, even when there is no AdS-space at all. Seing that the $U(1)$ isometry 
of Kerr is enhanced to $SL(2,\RR)$ and further to {\it Virasoro}, it is possible that 
the $U(1)\times U(1)$ isometry of general black 
holes might be enhanced to $SL(2,\RR)\times SL(2,\RR)$ and on to {\it Virasoro}$^2$. 
One tangible piece of evidence for this structure is the remarkable quantization rule 
\refs{\LarsenGE,\CveticUW}
\eqn\xb{
{1\over (8\pi G_4)^2}A_+ A_- ={\rm integer}~,
}
satisfied by the outer/inner horizon area in an astonishing variety of examples. We 
will not pursue this wider perspective further in this paper but it is clearly one
of the underlying motivations.

This paper is organized as follows. In section 2 we review the computation of greybody
factors for 4D black holes with charges. We emphasize the verification of 
matching condition for rotating black holes. In section 3 we turn to the greybody factors
for 5D black holes, with charges and two independent angular momenta. In section
4 we discuss the CFT interpretation of these scattering amplitudes. Finally, in 
section 5, we situate Kerr/CFT relative to the CFTs describing more general 
black holes. In particular we relate the temperatures that appear naturally in the
greybody factors to the Frolov-Thorne temperature employed in Kerr/CFT.

\newsec{Greybody Factors for 4D Rotating Black Holes}
In this section we review the Klein-Gordon equation in the background of
the general 4D rotating black holes with charges. We discuss the matching procedure
that leads to its solution, with emphasis on the extreme rotating limit.

\subsec{The Wave Equation}
We consider the general asymptotically flat 4D black hole with rotation, and also
four independent $U(1)$ charges. The solution was constructed in \CveticKV. Following
\CveticXV\ we present the massless Klein-Gordon equation as
\eqn\aa{\eqalign{&
\Big[ 4 {\partial\over\partial x}(x^2 - {1\over 4}){\partial\over\partial x}
+  
{1\over x-{1\over 2}}\left( {\omega\over\kappa_+} - m{\Omega\over\kappa_+}\right)^2
- {1\over x+{1\over 2}}\left( {\omega\over\kappa_-} - m{\Omega\over\kappa_+}\right)^2 +  4 \tilde{j}(\tilde{j}+1) \cr
& ~~~~~~~~~~~~~ + 8G_4 xM\Delta \omega^2+  x^2 \Delta^2 \omega^2  \Big]\Phi=0~.
}}
We employ the radial coordinate 
\eqn\aaa{
x = {r - {1\over 2}(r_+ + r_-)\over r_+ - r_-}~,
}
which is designed so that the two horizons
\eqn\aab{
r_\pm = {1\over 4} ( \mu \pm \sqrt{\mu^2 - l^2})~,
}
are at $x=\pm {1\over 2}$. The overall scale of the black hole is set by $r_+ + r_- = {1\over 2} \mu$.
The departure from extremality is encoded in 
\eqn\aac{
\Delta = 2 (r_+ - r_-)= \sqrt{\mu^2-l^2}~.
}
We have assumed that the dependence of the wave function on the temporal and angular
Killing vectors is 
\eqn\aaca{
\Phi\propto e^{-i\omega t'+im\phi'}~,
}
and we replaced the derivatives
$\partial_{t'}$ and $\partial_{\phi'}$ in the Laplacian accordingly. 

The dependence of the wave function on the polar coordinate $\theta$ is 
determined by the angular operator
\eqn\aad{
\tilde{\Lambda} = - {1\over\sin\theta}{\partial\over\partial\theta}\sin\theta{\partial\over\partial\theta} +{m^2\over\sin^2\theta} - {1\over 16}l^2\omega^2\cos^2\theta- {1\over 16}\mu^2\omega^2\left( 1 + \sum_{i<j}\cosh 2\delta_i \cosh 2\delta_j\right)~,
}
For the purposes of the radial equation we can think of this operator as a constant
(its eigenvalue)\foot{In \CveticUW\ we used the notation 
$
\zeta = {1\over 2} + \tilde {j} $. 
The recent work \BredbergPV\ similarly used $\beta = {1\over 2} + \tilde {j}$.}:
\eqn\aae{
\tilde{\Lambda} \to \tilde{j}(\tilde{j}+1)~.
}
In the special case of low energy $\mu\omega\ll \tilde{j}+{1\over 2}$ (which for near
extreme Kerr implies 
also $l\omega\ll \tilde{j} + {1\over 2}$), the angular wave function is just a spherical 
harmonic with angular momentum $j=\tilde{j}$. We will in fact not assume
low energy and so the generalized angular momentum $\tilde{j}$ is just 
a separation constant defined through \aae.  It takes on a sequence of 
discrete values that are not necessarily integral \foot{Although the eigenvalue 
$\tilde{j}(\tilde{j}+1)$ must be real, there are parameters for which $\tilde{j}$ becomes 
complex. In \BredbergPV\ this possibility was interpreted as a genuine instability,
interpreted in bulk as Schwinger pair production \refs{\PiolinePF,\KimXV}. Our computation
applies only when $\tilde{j}$ is real.}. 

\subsec{Parametric Representation of Black Hole Variables}
We use a parametric form for the physical variables of the black holes
\eqn\ba{\eqalign{
8G_4 M & = {1\over 2} \mu\sum_{i=1}^4\cosh 2\delta_i~,\cr
8G_4 Q_i & = {1\over 2} \mu\sinh 2\delta_i~,\cr
8G_4 J & = {1\over 2}\mu l \left( \prod_{i=1}^4\cosh\delta_i  - \prod_{i=1}^4\sinh\delta_i \right)~.
}}
The variables $\mu, l$ have dimension of length. Our charges $Q_i$ have dimension of
mass while the angular momentum $J$ is dimensionless. The special case of Kerr-Newman
black holes corresponds to having just one charge $Q\equiv {1\over 2}Q_i$ (for any $i$).  

The surface accelerations  $\kappa_\pm$ of the outer and inner horizons are encoded in
the R- and L-temperatures $T_{R,L}=\beta_{R,L}^{-1}$ with the parametric form 
\eqn\bb{\eqalign{
\beta_R &= {2\pi\over\kappa_+} + {2\pi\over\kappa_-} = {2\pi\mu^2\over\sqrt{\mu^2-l^2}}\left(\prod_i\cosh\delta_i+\prod_i\sinh\delta_i\right)~,\cr
\beta_L &= {2\pi\over\kappa_+} - {2\pi\over\kappa_-} = 2\pi\mu\left(\prod_i\cosh\delta_i-\prod_i\sinh\delta_i\right)~.
}}
The (inverse) Hawking temperature are given terms of these as
\eqn\bbb{
T_H^{-1} = \beta_H = {2\pi\over\kappa_+} = {1\over 2} ( \beta_R + \beta_L)~.
}
The angular velocity is parametrized as
\eqn\bba{
{1\over\kappa_+}\Omega = {l\over\sqrt{\mu^2- l^2}}~.
}
For easy reference we also record two equivalent expressions for the black hole entropy
\eqn\bcb{\eqalign{
S &= 2\pi \left[ {1\over 16G_4} \mu^2 \left(\prod_i \cosh\delta_i + \prod\sinh\delta_i\right)
+ \sqrt{ {1\over 256G^2_4} \mu^4 \left(\prod_i \cosh\delta_i - \prod\sinh\delta_i\right)^2- J^2}\right]\cr
&= {2\pi\over 8G_4}\left[ {1\over 2} \mu^2 \left(\prod_i \cosh\delta_i + \prod\sinh\delta_i\right)
+ {1\over 2}\mu\sqrt{\mu^2-l^2}\left(\prod_i \cosh\delta_i - \prod\sinh\delta_i\right)\right]~.
}}

\subsec{Solving the Wave Equation}
We solve the radial equation \aa\ one region at a time, and then patch the partial
solutions together for the complete wave function.  

The near horizon region of the black hole involves all terms in \aa\ except those linear 
and quadratic in the radial coordinate $x$. The solution to this part of the equation
is essentially a hypergeometric function\foot{We have taken the azimuthal quantum
number $m$ into account by using the replacements 
$\beta_{R}\omega/2\to \beta_{R}\omega/2 - \beta_H m \Omega$, 
$\beta_H\omega\to \beta_H(\omega-m\Omega)$ noted in footnote 8 of \CveticXV.}
\eqn\ba{\eqalign{
\Phi_{\rm NH} &= \left( {x-{1\over 2}\over x+{1\over 2}}\right)^{-i{\beta_H\over 4\pi}(\omega-m\Omega)}
(x+{1\over 2})^{-1-\tilde{j}}
 \cr
& \times
F\left( 1 + \tilde{j} - {i\over 2\pi} (\beta_R {\omega\over 2} - \beta_Hm\Omega),1+\tilde{j} 
- {i\beta_L\omega\over 4\pi} 
, 1- i {\beta_H\over 2\pi}(\omega-m\Omega), {x-{1\over 2}\over x +{1\over 2}}\right)~.
}}
The complex conjugate expression is a linearly independent solution.

The asymptotic region of the black hole involves just the terms that are constant or
increase as a function of the radial coordinate $x$. The solution to this part of the 
equation alone is essentially Kummer's function, as usual for scattering with on a
potential with a long range force ($1/r$-component) and a 
centrifugal barrier ($1/r^2$-component). 

In favorable cases there is a ``matching'' region where both the near horizon and 
the asymptotic approximation apply. In this region the radial equation involves 
just the generalized angular momentum barrier $\tilde{j}(\tilde{j}+1)$ and
the kinetic energy. Accordingly, the near horizon wave function \ba\ and 
the asymptotic wave function both take the same, simple form
\eqn\baa{
\Phi_{\rm matching} = a x^{\tilde{j} }~.
}
The complementary solution $\sim x^{-1-\tilde{j}}$ is negligible. Therefore the
coefficient $a$ appearing in each of the two partial solutions of the wave equation 
must be the same, leading to the complete wave function. 

In the case of absorption by the black hole, boundary conditions at the outer horizon 
are chosen such that there is no outgoing wave there. This determines the coefficient
$a$ in the matching region, and so the resulting fluxes in the asymptotic region. The 
result for the absorption cross-section found using the steps outlined above 
is \CveticXV\foot{The absorption cross-section is not just the ratio of fluxes at the horizon and 
asymptotically: there is also the overlap with a plane wave. For the numerical factor we 
use (without detailed justification) the standard result also when $\tilde{j}$ is not an integer.}
\eqn\bbg{\eqalign{
\sigma_{\rm abs}(\omega) & = {\pi(2\tilde{j}+1)\over\omega^2} \cdot 
{2\beta_H(\omega-m\Omega)\over \pi\Delta\omega}
\cdot \left| {({\rm ampl})_0\over ({\rm ampl})_\infty}\right|^2 \cr
& = {(\Delta\omega)^{1+2\tilde{j}}\over\omega^2} \sinh {\beta_H (\omega - m\Omega)\over 2}
\left| \Gamma( 1 + \tilde{j} - {i\beta_L\omega\over 4\pi})
\Gamma( 1 + \tilde{j} - {i\over 2\pi}(\beta_R{\omega\over 2}- \beta_H m\Omega)) \right|^2 
\cr
& \times { (2\tilde{j}+1)\over
\Gamma(2\tilde{j}+1)^2\Gamma(2\tilde{j}+2)^2}e^{2\pi G_4 M\omega}|\Gamma( 1 + \tilde{j} + 2iG_4M\omega)|^2~.
}}
As we have emphasized, the only assumption we have made in finding this expression 
is the existence
of a suitable matching region. 
Before turning to the analysis of this expression we therefore need to justify that assumption. 
That is what we turn to next.

\subsec{Matching Region in the Near Extreme Kerr Limit}

We are interested in the near extreme Kerr limit $\mu\sim\ell$ with the scale $\mu$ 
arbitrary:
\eqn\bap{{\mu\over\Delta}\equiv 
{\mu\over\sqrt{\mu^2-l^2}}\gg 1~.
}
For the purposes of estimates we take charge parameters $\delta_i\sim 1$. The 
estimates are valid also for $\delta_i=0$, but extreme limits that involve $\delta_i\gg 1$ 
along with \bap\ require additional considerations. In the near extreme limit the inverse
temperatures $\beta_R, \beta_H\gg\mu$, but $\beta_L\sim\mu$ and $\Omega\sim\mu^{-1}$. 

As discussed in the previous subsection, the matching region is ``far away'' from 
the near horizon perspective so we must require that the near horizon terms must be subleading 
there. Rewriting the near horizon expressions 
we find the conditions
\eqn\bin{{1\over{x^2-\textstyle{1\over 4}}}\left({\beta_H\over{2\pi}}(\omega-m\Omega)\right)^2
\, +\, {1\over{x+\textstyle{1\over 2}}}\left({{\beta_L\omega}\over{2\pi}}\right)\left({\beta_H\over{\pi}}(\omega-m\Omega)-{{\beta_L\omega}\over{2\pi}}\right)\, \ll {\tilde j}({\tilde j}+1)\, .
}
Thus for them to be subleading one requires:
\eqn\baq{\eqalign{
{1\over x^2} \cdot \beta^2_H(\omega-m\Omega)^2 &\ll \tilde{j}(\tilde{j}+1)~,\cr
{1\over x} (\beta_L\omega)\cdot |2\beta_H(\omega-m\Omega)-\beta_L\omega| &\ll \tilde{j}(\tilde{j}+1)~.
}}
However, the matching region should nevertheless be ``near the black hole'' from 
the point of view of the asymptotically flat region. Thus the terms in \aa\ that are 
linear or quadratic in $x$ should be negligible. This condition can 
be expressed as the inequalities
\eqn\baqa{\eqalign{
x^2 \cdot {\Delta^2\over \mu^2}  \mu^2\omega^2 \ll \tilde{j}(\tilde{j}+1)~, \cr
x \cdot {\Delta\over\mu} \mu^2\omega^2\ll  \tilde{j}(\tilde{j}+1)~.
}}
 %
%
We need to establish that there is a range of $x\gg 1$ satisfying both \baq\ and \baqa.

For $x\gg 1$ a sufficient condition for \baq\ to be satisfied is:
\eqn\bins{\eqalign{
&\beta_H (\omega - m\Omega)\sim  1\, , \  \cr  &\beta_L\omega\sim1\, .}}
The first requirement requires that we probe the energies, which  are natural for the co-rotating observer, and the second requirement further constrains the energy regime  to be of order  ${\cal O}(\mu^{-1})$ (recall $\beta_L={\cal O}(\mu)$).
Thus,  we consider
\eqn\bara{\eqalign{
\mu\omega \sim m\sim 1~. 
}}
Recalling that $\mu\Omega\sim 1$ in the extreme limit \bap\ we can do this 
with the difference $\omega-m\Omega$ tuned such that \bins\ is satisfied 
even though $\beta_H\gg\mu$. 

The energy and azimuthal quantum number of the scalar field contribute 
to the angular operator \aad. With these scales taken as \bara\ it is 
natural to assume the generalized angular momentum $\tilde{j}\sim 1$ as well.  

At this point we have specified the properties of the scalar field probe completely. It
is then a simple matter to verify that the matching conditions \baq-\baqa\ are solved in 
the range 
\eqn\basu{
1\ll x\ll {\mu\over\sqrt{\mu^2-l^2}}~.
}

\subsec{A Formal Decoupling Limit}
It is instructive to revisit the limits we consider by comparing with 
a formal decoupling in real space, generalizing the NHEK limit \BardeenPX\ to the near 
extreme limit with charges. 

Introducing 
\eqn\bja{
 \epsilon\lambda = {1\over 2}(r_+ - r_-) ={1\over 4}\sqrt{\mu^2-l^2}~,
}
the near extreme limit is defined as $\lambda\to 0$ with the excitation scale
$\epsilon$ kept fixed along with the horizon scale $ {1\over 2} ( r_+ + r_-) = {1\over 2}\mu$. 

We focus on the near horizon region by introducing the scaling coordinate $U$ through
\eqn\bjb{
r = {1\over 2} ( r_+ + r_-) + \lambda U~.
}
We keep $U$ fixed as $\lambda\to 0$. 
The dimensionless coordinate $x = U/2\epsilon$ 
introduced in \aaa\ also remains a good coordinate in the scalling limit. 

In the wave equation \aa\ we assumed the form \aaca\ for the wave function. In other
words, we made the replacements
\eqn\bjc{\eqalign{
{\partial\over\partial t'}  & \to - i \omega~,\cr
{\partial\over\partial\phi'} & \to i m~. 
}}
Note that the asymptotic coordinates are defined with a prime from the outset. The near 
horizon observer more naturally employ the comoving coordinates
\eqn\bjd{\eqalign{
\cr
t & = \lambda t'~, \cr
\phi & = \phi'-  \Omega t'~.
}}
In these coordinates
\eqn\bjd{
\omega- m\Omega \to i (\partial_{t'}+\Omega\partial_{\phi'}) = i\lambda\partial_t \to \lambda 
\omega_{\rm com}~. 
}
The comoving energy $\omega_{\rm com}$ is kept finite in the scaling limit. 

The surface accelerations $\kappa_\pm\sim\lambda$ so all the terms in the first line
of the wave equation \aa\ remain finite in the scaling limit. In contrast, the asymptotic 
terms in the second line of \aa\ scale to zero. The scaling limit thus isolates the
near horizon region, including the matching region, while the asymptotically flat region 
decouples as $\lambda\to 0$. 
It is therefore sensible to propose a theory that controls the near horizon region alone.

\subsec{Superradiance}
At this point we have established that the existence of a suitable matching region in 
the case of extremal Kerr \bap. This means the absorption cross-section \bbg\ 
applies in this case. 

The non-trivial frequency dependence 
in the absorption cross-scetion \bbg\ all remains in the limit discussed in the 
previous two subsections: there is no further simplification beyond the one due 
to the existence of a matching region. 

In the formal decoupling limit $\lambda\to 0$ the absorption cross-section in fact
vanishes, because the overall prefactor scales to zero. Thus the black hole
does not absorb incoming waves, nor does it emit particles. This limit is therefore
truly a decoupling limit. 

It is interesting and surprising that the absorption cross-section \bbg\
may be negative
\eqn\caa{
\sigma_{\rm ab} ( \omega<m\Omega) < 0~.
}
In this situation the amplitude of the wave reflecting from the black hole is larger
than the incoming wave. This phenomenon is known as 
super-radiance \refs{\MisnerKX,\TeukolskyHA,\PressZZ}.

It is interesting to trace the origin of super-radiance in our set-up. 
Very near the (outer) horizon at $x\sim{1\over 2}$ the radial wave function \ba\ 
reduces to
\eqn\cab{
\Phi_{\rm NH} (x -  {1\over 2}\ll 1) \sim (x-{1\over 2})^{-i{\beta_H\over 4\pi}(\omega-m\Omega)}~.
}
An exponent with negative imaginary part corresponds to an incoming wave, as one 
expects for absorption by the black hole. However, for $\omega<m\Omega$ the 
exponent has positive imaginary part. Then the flux is flowing out from the horizon
so that, at infinity, more flux is reflected than is send in. 

\subsec{The Emission Spectrum}
The spectrum of emitted Hawking radiation follows from the
absorption cross-section by detailed balance. It becomes
\eqn\bbx{\eqalign{
\Gamma_{\rm em}(\omega) &= \sigma_{\rm abs}(\omega) {1\over e^{\beta_H(\omega-m\Omega) -1 } }
{d^3 k\over (2\pi)^3}\cr 
& = {(\Delta\omega)^{1+2\tilde{j}}\over 2\omega^2} e^{\beta_H(\omega-m\Omega)/2}
\left| \Gamma( 1 + \tilde{j} - {i\beta_L\omega\over 4\pi})
\Gamma( 1 + \tilde{j} - {i\over 2\pi}(\beta_R{\omega\over 2}- \beta_H m\Omega)) \right|^2 
\cr
& \times { (2\tilde{j}+1)\over
\Gamma(2\tilde{j}+1)^2\Gamma(2\tilde{j}+2)^2}e^{2\pi G_4 M\omega}|\Gamma( 1 + \tilde{j} + 2iG_4M\omega)|^2 {d^3k\over (2\pi)^3}~.
}}
The emission spectrum does not exhibit superradiance: superradiance is stimulated
emission so it relies on the incoming quanta as well. However, it is more convenient for
the discussion of the CFT description. 

The frequency dependence in the final line of \bbx\ is due to the long range nature of
the interaction. This term is present for all processes involving $1/r$ forces, including
atomic and nuclear scattering. Although it arises from a hypergeometric function it can 
presumably not be interpreted as due to an underlying CFT. In section 4 we will therefore 
seek to understand just the frequency dependence in the first line of \bbx. 

\newsec{Near-Extreme Kerr Black Holes in D=5 }
In this section we carry out the analysis of near extreme Kerr black holes in five
dimensions. We maintain all three $U(1)$ charges and two independent angular momenta. 

\subsec{The Scalar Wave Equation}
The asymptotically flat black hole solution in 5D with independent values for
the two angular momenta and also three independent charges was found in \CveticXZ. 
The corresponding Klein-Gordon equation was presented in \CveticUW\ as
\eqn\ga{\eqalign{&
\left[ 4{\partial\over\partial x}(x^2 - {1\over 4}){\partial\over\partial x}
+ 
{1\over x-{1\over 2}}\left( {\omega\over\kappa_+} - m_R{\Omega_R\over\kappa_+} - 
m_L{\Omega_L\over\kappa_+}\right)^2
- {1\over x+{1\over 2}}\left( {\omega\over\kappa_-} - m_R{\Omega_R\over\kappa_+} +
m_L{\Omega_L\over\kappa_+}\right)^2\right. \cr
& \left.    -  \tilde{j}(\tilde{j}+2) +  x\Delta\omega^2\right]\Phi_0=0~.
}}
The radial coordinate
\eqn\gaa{
x = {r^2 - {1\over 2}(r^2_+ + r^2_-)\over r^2_+ - r^2_-}~,
}
is designed to put the horizons 
\eqn\gab{
r^2_\pm = {1\over 2} \left( \mu \pm \sqrt{ (\mu-(l_1-l_2)^2) (\mu-(l_1+l_2)^2)}\right)~,
}
at $x=\pm {1\over 2}$ for all values of the black hole parameters. 
The departure from extremality (which may be arbitrary at this point) is encoded in
\eqn\gac{
\Delta = r^2_+ - r^2_-=\sqrt{ (\mu-(l_1-l_2)^2) (\mu-(l_1+l_2)^2)}~.
}

The full angular Laplacian for the problem is  
\eqn\gae{\eqalign{
{\hat\Lambda} =  - {1\over\sin 2\theta}{\partial\over\partial\theta} \sin2\theta {\partial\over\partial\theta}
- {1\over\sin^2\theta} {\partial^2\over\partial\phi^2} - {1\over\cos^2\theta} {\partial^2\over\partial\psi^2}
+ (l_1^2+l_2^2)\omega^2 + (l_2^2-l_1^2)\omega^2\cos 2\theta - M\omega^2~. 
}}
We denote the eigenvalue of this operator $\tilde{j}(\tilde{j}+2)$. Accordingly, we inserted this value 
of the angular momentum barrier in the radial equation \ga. 
At low energy $l_{1,2}\omega^2, M\omega^2\ll 1$ our notation $\tilde{j}$ reduces to the usual
angular momentum $j$, which in that limit labels the quadratic Casimirs of the
rotation group $SO(4)\simeq SU(2)\times SU2)$. However, we will not assume 
that the energy is small and so the generalized angular momentum $\tilde{j}$ 
is just a notation for the separation constant of the Klein-Gordon equation.\foot{As in 4D, 
$\tilde{j}(\tilde{j}+2)$ must be real, but in general this combination can be 
less than $-1$ so that $\tilde{j}$ may acquire an imaginary part.}

\subsec{Parametric form of Black Hole Variables}
In the general case with three $U(1)$ charges it is essential that we employ the
parametric representation of black hole variables \CveticXZ
\eqn\gba{\eqalign{
{4G_5\over\pi}M &= {1\over 2} \mu\sum_{i=1}^3\cosh 2\delta_i~,\cr
{4G_5\over\pi}Q_i & = {1\over 2} \mu \sinh 2\delta_i~,~~~ (i=1,2,3)~,\cr
{4G_5\over\pi}J_{R,L} & = {1\over 2} \mu (l_1\pm l_2)
\left( \prod_{i=1}^3\cosh\delta_i \mp \prod_{i=1}^3 \sinh\delta_i\right)~.
}}
Note that in 5D the scale $\mu$ has dimension of length {\it squared}. The parametric
angular momenta $l_{1,2}$ are lengths and the parametric charges are $\delta_i$
dimensionless. \foot{We indicate Newton's constant explicitly. For the value 
$G_5={\pi\over 4}$ the formulae simplify and $Q_i$ becomes integral in the
simplest string theory embedding (see eg. \LarsenXM).}

The surface accelerations $\kappa_\pm$ in the radial equation \ga\ are equivalent to
the inverse temperatures
\eqn\gb{
\beta_{R,L} = {2\pi\over\kappa_+}\pm {2\pi\over\kappa_-}~,
}
which in turn have the parametric form
\eqn\gc{\eqalign{
\beta_L &= {2\pi\mu\over
\sqrt{\mu-(l_1-l_2)^2}} ( \prod_i \cosh\delta_i -\prod_i \sinh\delta_i)~,\cr
\beta_R &= {2\pi\mu\over
\sqrt{\mu-(l_1+l_2)^2}}(\prod_i \cosh\delta_i +\prod_i \sinh\delta_i)~.
}}
and the inverse Hawking temperature  $\beta_H={{2\pi}\over \kappa_+}$. 

The angular velocities in the radial equation \ga\ have the parametric forms
\eqn\gd{\eqalign{
\beta_H\Omega_L &
={2\pi (l_1-l_2)\over\sqrt{\mu-(l_1-l_2)^2}}~,\cr
\beta_H\Omega_R &
={2\pi (l_1+l_2)\over\sqrt{\mu-(l_1+l_2)^2}}~.
}}

For later reference we also record the black hole
entropy
\eqn\gda{\eqalign{
S &= 2\pi\sqrt{ {\pi^2\over 64G^2_5} \mu^3 \left(\prod_i\cosh\delta_i + \prod\sinh\delta_i\right)^2-J^2_L}+
2\pi\sqrt{ {\pi^2\over 64G^2_5}  \mu^3 \left(\prod_i\cosh\delta_i - \prod\sinh\delta_i\right)^2-J^2_R}~.
}}

\subsec{Wave Functions and Greybody Factors}
The radial equation \ga\ cannot be solved analytically in general. However, in the
near horizon region where the term linear in $x$ can be neglected the equation is
hypergeometric with solution\CveticUW\foot{The notation of \CveticUW\ is $\xi = 1+{1\over 2}\tilde{j}$.}
\eqn\gee{\eqalign{
\Phi_{\rm NH} (x)
& = \left( {x-{1\over 2}\over x+{1\over 2}} \right)^{-i{\beta_H\over4\pi}(\omega-m_L\Omega_L - m_R\Omega_R)}
(x+{1\over 2})^{ - 1-{1\over 2}\tilde{j} } \times 
F\left( 1 +{1\over 2}\tilde{j} -{i\over 2\pi} ({\beta_R\omega\over 2} - \beta_Hm_R\Omega_R),\right. \cr
& \left. 1+{1\over 2}\tilde{j} -{i\over 2\pi} ({\beta_L\omega\over 2} -  \beta_Hm_L\Omega_L),
1-i{\beta_H\over 2\pi}(
\omega - m_L\Omega_L- m_R\Omega_R), {x-{1\over 2}\over x+{1\over 2}}\right)~.
}}
The wave function was chosen with incoming boundary conditions. The complex
conjugate wave function is a linearly independent solution, with outgoing boundary
condition. 
The asymptotic behavior of \gee\ for large $x$ takes the form
\eqn\gea{
\Phi_{\rm NH} (x) \sim a x^{{1\over 2}\tilde{j}}~.
}
The solution in the asymptotic region where the horizon terms with singularities as $x=\pm {1\over 2}$
can be neglected is also simple: it is just a Bessel function. In the short distance limit this asymptotic
wave function takes the same form as \gea. In cases where an overlapping regime of applicability
of the two regimes can be established the coefficient $a$ for the two regional wave functions
must agree, and then the full wave function follows. Comparing the asymptotic flux to the
one at the horizon, we find the transmission coefficient
\eqn\gfa{\eqalign{
|T_{\tilde j}|^2 &= \beta_H(\omega - m_L\Omega_L - m_R\Omega_R)
\left( {\sqrt{\Delta}\omega\over 2}\right)^{2+2\tilde{j}}\cr &~~~\times
\left| { \Gamma( 1 + {1\over 2} \tilde{j} - {i\over 2\pi} (\beta_L {\omega\over 2} - \beta_H m_L\Omega_L))
\Gamma( 1+ {1\over 2} \tilde{j} - {i\over 2\pi} (\beta_R {\omega\over 2} - \beta_H m_R\Omega_R))
\over \Gamma(\tilde{j})\Gamma(1+\tilde{j})
\Gamma(1 - {i\over 2\pi}\beta_H (\omega - m_L \Omega_L  - m_R \Omega_R)) }\right|^2~.
}}
Expanding one of the $\Gamma$-functions, the absorption cross-section 
becomes\foot{The transmission coefficient and the cross-section are related by the 
overlap between our wave function in spherical coordinates
and a plane wave. This is difficult to compute because the solutions to the angular equation \gae\ 
are involved when there is no spherical symmetry. Guided by spherical symmetry, we 
use the overlap ${4\pi(\tilde{j}+1)^2/\omega^3}$, knowing that this expression should receive
small corrections.} 
\eqn\gf{\eqalign{
\sigma_{\rm abs}(\omega) &= {8\pi\over\omega^3}\sinh\left( {1\over 2}\beta_H(\omega - m_L\Omega_L - m_R\Omega_R)\right)
\left( {\sqrt{\Delta}\omega\over 2}\right)^{2+2\tilde{j}} 
{(\tilde{j}+1)^2\over |\Gamma(\tilde{j})\Gamma(1+\tilde{j})|^2} \cr &~~~\times
\left|  \Gamma( 1 + {1\over 2} \tilde{j} - {i\over 2\pi} (\beta_L {\omega\over 2} - \beta_H m_L\Omega_L))
\Gamma( 1+ {1\over 2} \tilde{j} - {i\over 2\pi} (\beta_R {\omega\over 2} - \beta_H m_R\Omega_R))
\right|^2~.
}}

\subsec{The Near Extreme Limit and Matching Conditions}

The 5D near extreme Kerr limit takes one of the two angular momenta large,
keeping the other at moderate values. Without loss of generality, we take 
$J_R\sim J_{R,{\rm max}}$, with $J_L$ arbitrary. In our parametric notation we take 
\eqn\ha{
{\sqrt{\mu}\over {\sqrt{\mu - (l_1+l_2)^2}}} \gg 1~.
}
The variables are otherwise not constrained so we can estimate 
$\mu-(l_1 - l_2)^2\sim \mu$ for the combination that controls the other angular momentum ($J_L$).
The non-extremality parameter \gac\ becomes 
\eqn\hax{
\Delta\sim 2\sqrt{l_1l_2} \sqrt{\mu - (l_1+l_2)^2} \le \sqrt{\mu} \sqrt{\mu - (l_1+l_2)^2}\ll \mu
}
in the limit \ha.   In our limit $\beta_H\sim\beta_R\gg \beta_L$ .
Also, $\Omega_R\sim\mu^{-1/2}$ and $\Omega_L\sim\beta_H^{-1}$ 
so $\Omega_R\gg \Omega_L$. As in 4D we take the charge 
parameters $\delta_i\sim 1$  in our estimates. 

The matching region is a range of $x$ where the angular momentum
barrier dominates the near horizon terms.  Rewriting the near horizon 
expressions we find the conditions:
\eqn\binf{\eqalign{&{1\over{x^2-\textstyle{1\over 4}}}\left({\beta_H\over{2\pi}}(\omega-m_R\Omega_R-m_L\Omega_L)\right)^2
\, +\, \cr 
&{1\over{x+\textstyle{1\over 2}}}\left({{\beta_L\omega}\over{2\pi}}-{{\beta_Hm_L\Omega_L}\over{\pi}}\right)\left({\beta_H\over{\pi}}(\omega-m_R\Omega_R-m_L\Omega_L)-({{\beta_L\omega}\over{2\pi}}-{{\beta_Hm_L\Omega_L}\over{\pi}})\right) \ll {\tilde j}({\tilde j}+2)\, .
}}
Since $\Omega_L\ll \Omega_R$  and ${\tilde j}\sim 1$,  sufficient conditions for these terms to be subleading  for   $x\gg 1$ are: 
%
%
%
\eqn\dcp{\eqalign{
&\beta_H(\omega-m_R\Omega_R) \sim 1~,\cr
&\beta_L\omega- \beta_Hm_L \Omega_L\sim 1\, ~. 
}}
The first condition is the most delicate since $\beta_H$ is large. We satisfy it by focussing on modes
with their natural energy and azimuthal quantum number, but a cancellation so that \dcp\ is
satisfied. The second condition is almost automatic since neither 
$\beta_L$ or $\beta_H\Omega^L$ is large in the near extreme limit. In formulae, we
take:
\eqn\hd{
\sqrt{\mu} \omega\sim m_R\sim m_L \sim 1~,
}
with the precise values of $\omega, m_R$ tuned so that \dcp\ remains satisfied even though
$\beta_H\gg\sqrt{\mu}$ in the limit \ha.

In the matching region the angular momentum barrier must also dominate the term 
encoding asymptotic Minkowski space, ie. the term in \ga\ that is linear in $x$. This
gives the condition
%
%
\eqn\dcc{x{\Delta\over\mu}(\mu\omega^2)\ll \tilde{j}(\tilde{j}+2)~.}
%
 The natural magnitude for the generalized
angular momentum is similarly $\tilde{j}\sim 1$, since the expression \gae\ receives
contributions from terms of the order \hd. 

We can now verify that the conditions \dcp\ -\hd\ on the scalar wave are sufficient 
to satisfy the matching conditions \binf\  and  \dcc\ in the range
\eqn\he{
1\ll x \ll \sqrt{\mu\over \mu - (l_1+l_2)^2} ~.
}
This is what is needed to justify the greybody factors \gf.

As in 4D (section 2.5) we could formalize the estimates in this section such that
the validity of the approximations are recast as a formal limit, rendering the
near horizon region (including the matching region) properly decoupled
from the asymptotically flat space. We will mostly refer to the approximate notation
detailed in this section. 

\subsec{Superradiant Greybody Factors}
At this point we have justified the use of the matching procedure for the 5D extreme
black holes with charge. All the structure in the absorption cross-section \gf\ persists
in the scaling limit, there are no further simplifications. 

As in 4D, the absorption cross-section may turn negative, corresponding to superradiance.
The condition for this phenomenon is
\eqn\hg{
\beta_H ( \omega - m_L \Omega_L - m_R \Omega_R)
= (\beta_L {\omega\over 2} - \beta_H m_L \Omega_L)
+(\beta_R {\omega\over 2} - \beta_H m_R \Omega_R)<0~.
}
The two parenthesis in the last expression are both of order $1$ in our scaling limit.
Superradiance can therefore be realized in non-trivial ways in 5D: it can be due to
level inversion in {\it either} the L {\it or} the R side. 

\subsec{The Emission Spectrum}
We may recast the absorption cross-section \gf\ as an emission amplitude for Hawking
radiation, using detailed balance. The result is 
\eqn\hi{\eqalign{
\Gamma_{\rm em} (\omega) &= \sigma_{\rm abs}(\omega) 
{1\over e^{\beta_H( \omega - m_L \Omega_L - m_R \Omega_R)}-1} {d^4k\over (2\pi)^4}\cr
&= {4\pi\over\omega^3}
\left( {\sqrt{\Delta}\omega\over 2}\right)^{2+2\tilde{j}} {(\tilde{j}+1)^2\over |\Gamma(\tilde{j})\Gamma(1+\tilde{j})|^2}
e^{-{1\over 2}\beta_H(\omega - m_L\Omega_L - m_R\Omega_R)}
 \cr &\times
\left|  \Gamma( 1 + {1\over 2} \tilde{j} - {i\over 2\pi} (\beta_L {\omega\over 2} - \beta_H m_L\Omega_L))
\Gamma( 1+ {1\over 2} \tilde{j} - {i\over 2\pi} (\beta_R {\omega\over 2} - \beta_H m_R\Omega_R))
\right|^2{d^4k\over (2\pi)^4}~.
}}
In the case where $U(1)$ charges and two independent angular momenta are included
the four potentials $\beta_{R,L} and \beta_H\Omega_{R,L}$ are independent. This gives
significant structure to the amplitude \hi.

We have maintained the notation appropriate for the asymptotic observer. However, in the
near horizon theory it is more natural to introduce the rescaled potential 
$\tilde{\beta_H}=\lambda\beta_H$ and the corresponding comoving energy
$\omega_{\rm com} = \lambda (\omega - m_R\Omega_R)$ with the scaling parameter 
$\lambda\sim {\mu\over \sqrt{\mu^2-(l_1+ l_2)^2}}$ taken to be small. The rescaled
quantities are finite even in the formal scaling limit $\lambda\to 0$.

\newsec{The CFT Model}

In this section we model the emission amplitudes from a microscopic point of view. 
The presentation follows our previous papers 
\refs{\CveticUW\CveticXV\CveticVP}, now adapted to the Kerr/CFT
context. In comparison with the recent work \BredbergPV\ we include all the overall
frequency dependent factors. We also keep all four $U(1)$ charges in the 4D theory, 
and we include both angular momenta in the 5D theory. These additional black hole 
parameters makes the general structure more transparent and makes the relation to 
the BPS cases clearer.  We first consider the 5D theory, and then briefly the 4D case. 

\subsec{5D Emission Spectrum from CFT}

The working assumption of the microscopic model is that the entire near horizon region, 
including the matching region, can be described by a dual CFT, generalizing 
the 4D Kerr/CFT \GuicaMU.

That the near horizon region should be dual to some quantum field theory is suggested 
by the decoupling of this region from the asymptotically flat space. That the theory 
should be a CFT is made possible by the classical fields reducing to 
hypergeometric functions, which are the characters of the $SL(2,\RR)$ group. 
The geometrical origin of the $SL(2,\RR)$ is the isometry group the AdS$_2$ factor in the 
geometry, and the wave equation is the $SL(2,\RR)$ Casimir.

In the description where the near horizon region is replaced by a CFT, the emission
of quanta embodied in \hi\ is due to couplings 
\eqn\jaa{\Phi_{\rm bulk}{\cal O}^{(h,{\bar h})}
}
between bulk modes $\Phi_{\rm bulk}$ and operators ${\cal O}^{(h,{\bar h})}$ in the CFT. 
The structure of the resulting emission depends primarily on the conformal weights 
$(h,{\bar h})$ of the operator.  The value of the conformal weight 
\eqn\ja{
h = {\bar h} = 1+ {1\over 2}\tilde{j}~,
}
can be read off from the asymptotic behavior \gea\ near the boundary of the near horizon 
region. 

In situations where the black hole background has spherical symmetry the difference
$h-{\bar h}$ measures the bulk spin $s$ and so it is obvious that $h={\bar h}$ for scalar 
fields in bulk (see eg. \refs{\MaldacenaIH,\deBoerIP}). 
The Kerr black hole is not spherically symmetric and so $h={\bar h}$ is not clear
{\it a priori} \BredbergPV. 

The canonical thermal two-point function of chiral operator with conformal weight $h$ is
specified by the singularity $\sim z^{-2h}$ and the periodicty $2\pi\beta^{-1}$: 
\eqn\jb{
G^h_\beta(z) = \left( {\pi/\beta}\over \sinh\left( {\pi z/\beta}\right) \right)^{2h}
}
The Fourier transform is
\eqn\jc{
G^h_\beta \left({\omega\over 2}\right) = 
\left( {2\pi\over\beta}\right)^{2h-1} e^{-\beta\omega/4}
{1\over\Gamma(2h)} \left| \Gamma(h + {i\beta\omega\over 4\pi}) \right|^2
}
The two point function of an operator ${\cal O}^{(h,{\bar h})}$ with conformal weights \ja\ thus 
gives the contributions to the emission amplitude from the CFT operators: 
\eqn\jd{\eqalign{
\Gamma_{\rm em}(\omega) & \propto \left({4\pi^2\over\beta_R\beta_L}\right)^{\tilde{j}+1} 
e^{-\beta_L(\omega-m_L\Omega_L)/4-\beta_R(\omega-m_R\Omega_R)/4}\cr
& \times \left| \Gamma(1+{1\over 2}\tilde{j} + {i\over 2\pi}(\beta_R{\omega\over 2}-\beta_H m_R\Omega_R)) \right|^2\cdot
\left| \Gamma(1+{1\over 2}\tilde{j} + {i\over 2\pi}(\beta_L{\omega\over 2}-\beta_H m_L\Omega_L)) \right|^2}}
This goes a long way towards accounting for the supergravity expression \hi. As explained 
after \hi, we maintain the notation appropriate for the asymptotic observer even though. Since the
CFT knows only about comoving energies and rescaled temperatures, it is those combinations
that appear in \jd. 

The details of the emission will depend on the coupling \jaa\
between the CFT operator and the bulk field. If \jaa\ is literally the coupling, the 
only additional frequency dependence is $\omega^{-1}$ from the standard 
normalization of the outgoing wave function $\Phi_{\rm bulk}$. 
However, generally the coupling must also include derivatives and numerical group 
theory factors (such as $\Gamma$-matrices) in order to ensure Lorentz invariance
and other symmetries. At low energy, the coupling to spin $j$ involves precisely $j$ 
derivatives that act on the outgoing wave function, giving a factor $\omega^{j}$ in the 
amplitude, and the square of that in the 
probability \refs{\MaldacenaIH,\MathurET,\GubserQR,\CveticAP}. 
In Kerr/CFT there is not enough known about the dual theory that we can construct the 
coupling to bulk fields in any detail. Nevertheless, it is reasonable to expect such 
couplings to lead to an overall frequency dependence 
\eqn\je{
\omega^{2\tilde{j}-1}~.
}
It is the ``far away'' frequency $\omega$ rather than either of 
the ``near horizon'' (comoving) frequencies $\omega-m_{L,R}\Omega_{L,R}$
that enter in this factor. The reason is that the (generalized) derivatives in the 
coupling can be taken to act on the bulk wave function $\Phi_{\rm bulk}$ which
only reaches into the matching region. Thus the coupling is sensitive to the 
deformation of the sphere due to rotation ($\tilde{j}$ rather than $j$) but not to 
the motion of the near horizon region. 

The emission rate \jd\ in the microscopic model, with the prefactor \je, should be compared 
with the supergravity result \hi. A useful relation is 
\eqn\jf{
{\beta_R\beta_L\Delta\over 2\pi} = {4G_5\over\pi} {\cal L}_5~,
}
where we have introduced the length scale ${\cal L}_5 $ through \refs{\CveticUW,\KastorGT}
\eqn\jg{
{4G_5\over\pi}{\cal L}_5 = 2\pi\mu^2
\left( \prod_{i=1}^3\cosh^2\delta_i - \prod_{i=1}^3\sinh^2\delta_i\right)~.
}
The two expressions depend identically on all black hole parameters. One formal
discrepancy arises because the CFT expressions like \jc\ were written by convention 
in units where the CFT is defined on a space of unit length. The comparison 
determines that length scale as ${\cal L}_5$ given in \jg. 

We do not have a derivation of this length scale from first principle in the present context.
However, in the near BPS limit $\delta_{1,2}\gg 1$ the length scale depends on just
two of the three charges, and on the length scale associated with the third charge. In
the standard $D1-D5-KK$ duality frame, the scale becomes
\eqn\jga{
{\cal L}_{5,{\rm BPS}} = 2\pi n_1 n_5 R~.
}
This is the ``long string scale'', corresponding to maximal winding around the
compact $KK$-circle \MaldacenaDS. It is this scale that controls emission amplitude
in many simpler contexts (see eg \refs{\DasWN,\MathurET,\MaldacenaIX,\CveticXH}). 
In the next section we
discuss the corresponding length scale for near extreme Kerr.

We have not attempted to reproduce the overall numerical factors in the emission 
amplitude from a microscopic point of view. In the simplest case of low energy emission 
from a spherical symmetric black hole the numerical factor was understood long time 
ago \DasWN. There is also (at least) a partial understanding of the numerical factors 
pertaining to higher partial waves \refs{\MaldacenaIH,\MathurET,\GubserQR} but those 
depend on the explicit coupling between bulk modes and the CFT which is not available 
here. 

The overall scaling of the amplitudes represents an interesting point. 
In the CFT amplitude \jd\ the overall normalization include $\beta^{-\tilde{j}-1}$.
It is the CFT temperature that enters, so it would be more correct to write 
$\tilde{\beta}^{-\tilde{j}-1}_R = \lambda^{-\tilde{j}-1}\beta^{-\tilde{j}-1}$, in the notation 
after \hi. In other words, the supergravity amplitude is suppressed in the scaling parameter 
$\lambda$, as one expects when the near horizon theory is fully decoupled; 
but the CFT amplitude is not suppressed, because everything is written in terms of rescaled 
variables. 

\subsec{4D Emission Spectrum from CFT}
The microscopic model that gives an interpretation of the 4D supergravity 
emission amplitude is very similar to the 5D model so we shall just summarize
the main formulae. 

The conformal weight of the CFT operator that is responsible for the emission can
be read off from the wave function \baa\ in the matching region \CveticXV
\eqn\ka{
h={\bar h} = 1 + \tilde{j}~.
}
The two point correlations function of the operator is again \jc\ in Fourier space. 
The normalization of the outgoing bulk wave function and the frequency dependence
from the couplings combine to give an overall frequency dependence
\eqn\kb{
\omega^{2\tilde{j}-1}~.
}
Collecting these factors give the emission amplitude
\eqn\kc{\eqalign{
\Gamma_{\rm em}(\omega) & \propto \left({4\pi^2\over\beta_R\beta_L}\right)^{2\tilde{j}+1} 
\omega^{2\tilde{j}-1} e^{-\beta_H(\omega-m\Omega)/2}\cr
& \times \left| \Gamma(1+\tilde{j} + {i\over 2\pi}(\beta_R{\omega\over 2}-\beta_H m\Omega)) \right|^2\cdot
\left| \Gamma(1+\tilde{j} + {i\over 4\pi}\beta_L\omega) \right|^2~.
}}
The complete dependence on the frequency $\omega$ and the azimuthal quantum number $m$ 
agrees precisely with \bbx\ (except for the Coulomb factors in the last line \bbx\ which should 
be neglected, as explained just after \bbx). 

The dependence on the black hole parameters can be compared by using the relation 
\eqn\kca{
{\beta_R\beta_L\Delta\over 2\pi} = 8G_4{\cal L}_4~,
}
where we have introduced the length scale ${\cal L}_4$ through \refs{\CveticUW,\KastorGT}
\eqn\kcb{
8G_4{\cal L}_4 = 2\pi\mu^3
\left( \prod_{i=1}^4\cosh^2\delta_i - \prod_{i=1}^4\sinh^2\delta_i\right)~.
}
The dependence on the black hole parameters also agrees except that, as in 5D, we 
have normalized our CFT correlation functions so that they depend on just the physical 
temperatures, but the size of the spatial circle has been scaled out. The comparison determines that length scale as \kcb. As in 5D we interpret this scale as the ``long string'' 
scale \refs{\GubserZP,\CveticJA}. 
In the limit where 
the black hole is nearly BPS there are three large charges $\delta_{1,2,3}\gg 1$ and \kcb\ 
reduces to 
\eqn\kf{
{\cal L}_{4,{\rm BPS}} = 2\pi n_1 n_2 n_3R~,
}
where $R$ is the size of a physical compactification circle. The general expression 
\kcb\ for the ``long string scale'' should be useful also away from the BPS limit. 
In the next section we make this expectation explicit in Kerr/CFT.

\newsec{Features of the Microscopic Theory}
In this section we extract some of the features of the microscopic theory.
We focus on the 4D theory for easy comparison with other works, and just
summarize the 
5D formulae. The important point as that we include all 
charges to appreciate the full structure.

\subsec{Phenomenological Model for General 4D Black Holes}
We start out very ambitiously, by writing the beginnings of a model for the {\it entire}
class of 4D black holes we consider, including black holes that are nowhere near
extremality. The working hypothesis is that all these black holes can be interpreted
as a 2D CFT in a periodic box with some unknown radius ${\cal R}_4$, and that the
entropy is captured by the standard high temperature expression:
\eqn\cc{
S = {\pi^2\over 3}  (c_L T_L +c_R T_R ) {\cal R}_4 ~.
}
The two temperatures $T_{L,R}$ we identify with the temperatures \bb\ that
appear in the greybody factors, and the entropy of the left and right movers independently
we take from the two terms in \bcb. These assumptions give expressions for the central 
charge in units of the box radius
\eqn\cd{
c_L {\cal R}_4 = c_R {\cal R}_4 = 
12\cdot {\mu^3\over 16G_4} \left( \prod_{i=1}^4  \cosh^2\delta_i - \prod_{i=1}^4\sinh^2\delta_i\right)~.
}
It is interesting that the central charge found this way is the same for the two chiralities. 

Let us now specialize to black holes that are extreme due to their rotation. In this
limit the entropy is exclusively due to the $L$-sector. It has been argued \refs{\GuicaMU} 
that in this situation the temperature of the $L$-sector is the Frolov-Thorne temperature \FrolovJH , 
which for general charges can be computed as
\eqn\cea{
\beta_{\rm FT} = {\partial \over\partial J} S(M=M_{\rm ext})= {2\pi J\over \sqrt{J^2 + {1\over 64G_4^2}\prod_{i=1}^4Q_i}}~.
}
This value is somewhat puzzling because it differs from the temperature $\beta_L$ in 
\bb\ which, as we have seen, appears quite prominently in the physical greybody 
factors, even in the extreme limit. To resolve this tension we note that the Frolov-Thorne
temperature defined in \cea\ is dimensionless. However, {\it the natural unit is the box radius},
so we can in fact identify the two proposed temperatures, after all: 
\eqn\ceb{
\beta_{\rm FT}=\beta_L/{\cal R}_4~.
}
Moreover, this identification determines the box size as
\eqn\cef{
{\cal R}_4 = \mu \left( \prod_{i=1}^4\cosh\delta_i + \prod_{i=1}^4\sinh\delta_i\right)~.
}
The CFT is of course scale invariant so ${\cal R}_4$ has no meaning in the microscopic 
theory: only the complex structure encoded in $\beta_{FT}$ makes sense. However,
the identification of observables at infinity involves ${\cal R}_4$. Additionally, 
${\cal R}_4\sim\tilde{\beta}_R$,
the rescaled temperature that does make sense in the CFT. 

At this point the central charge determined from \cd\ becomes
\eqn\ceg{
c_L = c_R = 12\cdot 
{\mu^2\over 16G_4} \left( \prod_{i=1}^4  \cosh\delta_i - \prod_{i=1}^4\sinh\delta_i\right) = 12J~.
}
The final equality followed from the relation between charges, mass and angular 
momentum in the extreme limit. The result for the central charge agrees with the
well known one from Kerr/CFT. In particular it does not depend on the value of the
$U(1)$ charges. This suggests that all the $U(1)$ charges are present in the
CFT from the outset. 

We are now ready to reconsider the length scale ${\cal L}_4$ that was extracted
from the greybody computations. Combining the formulae above, we find
\eqn\ceh{
{\cal L}_4 = J\cdot 2\pi{\cal R}_4~.
}
The effective length that appears in the scattering is therefore essentially the same
as the box size inferred from the simplest thermodynamic model. The only difference
is a rescaling related to the background angular momentum. This rescaling is 
reminiscent of the ``long string" rescaling \kf\ of BPS black holes. Our result
is a quantitative prediction for a similar phenomenon in Kerr/CFT.  

\subsec{The 5D model} 
In D=5 the entropy formula \gda\ can also quite generally be cast in the 
form \cc\ with two temperatures $T_{L,R}$  identified with those appearing in 
the greybody factors \gc. This procedure gives the central charges in units of 
the box radius ${\cal R}_5$
are
\eqn\ccd{
c_L {\cal R}_5 = c_R {\cal R}_5 = 12\cdot {{\pi \mu^2}\over 8G_5} \left( \prod_{i=1}^3  \cosh^2\delta_i - \prod_{i=1}^3\sinh^2\delta_i\right)~.
}
Again, the  central charges are  the same for  both chiralities. 

When specializing to the extreme  black holes, the entropy has only a contribution 
for the $L$-sector. The Frolov-Thorne temperature along the dominant (R) motion becomes:
\eqn\cce{\eqalign{
\beta_{\rm FT}&= {\partial \over\partial J_R} S(M=M_{\rm ext})= 
{2\pi J_R\over \sqrt{J^2_R - J^2_L + {4G_5\over\pi}\prod_{i=1}^3Q_i}}~.
}}
We can identify the greybody temperature $\beta_L$ with the Frolov-Thorne temperature
(in units of a box size) $\beta_{FT}$ by introducing the box-size
\eqn\cceb{
{\cal R}_5 = {\beta_L\over \beta_{\rm FT}}
= 2\pi\mu^{1/2} \left( \prod_{i=1}^3\cosh\delta_i +\prod_{i=1}^3\sinh\delta_i \right) ~,
}
As in 5D, the box size if of order the ``small'' temperature, in units if the scaling 
variable ${\cal R}_5\sim \tilde{\beta_R}$. 

The central charges determined from \ccd\ now become:
\eqn\cceg{\eqalign{
c_{L}& = c_{R} = 12\cdot {\mu^{3/2}\over 16G_5} 
\left( \prod_{i=1}^3\cosh\delta_i - \prod_{i=1}^3\sinh\delta_i\right) = 12J_R
}}
as expected. These expressions for the central charges are compatible with those found in 
\ChowDP\  where  Kerr/CFT techniques were employed \foot{Those 
corresponds to a more refined description where one Frolov-Thorne 
temperature for each angle is introduced $\beta_{{\rm FT}\phi},\beta_{{\rm FT}\psi}$. This gives
rise to two independent central charges, one for each component of the angular
momentum $J_{\phi,\psi}$. It is the sum of those that give \cceg.}.

Finally, we can now derive the  length scale ${\cal L}_5$ that was extracted
from the greybody computations. Combining the formulae above, we find
\eqn\cceh{
{\cal L}_5 = J_R \cdot 2\pi{\cal R}_5 ~.
}
Again the  rescaling is 
reminiscent of the ``long string" rescaling \kf\ of BPS black holes.

\bigskip
\noindent {\bf Acknowledgments:} \medskip \noindent
We thank A. Castro and A. Strominger
for discussions. MC thanks CERN for hospitality during
this work. FL thanks Aspen Center for Physics for hospitality while this work was completed. 
MC   is supported by the 
 DOE Grant DOE-EY-76-02- 
3071, the NSF RTG DMS Grant 0636606 and the Fay R. 
and Eugene L. Langberg Endowed Chair. 
FL is also supported by the DoE.

\listrefs
\end